\documentclass[aps,showpacs]{revtex4-1}
\usepackage{graphicx}
\usepackage{amsmath}
\usepackage{amsfonts}
\usepackage{amssymb}

\newcommand{\beq}{\begin{equation}}
\newcommand{\eeq}{\end{equation}}
\newcommand{\beqa}{\begin{eqnarray}}
\newcommand{\eeqa}{\end{eqnarray}}

\newcommand{\rr}{{\bf r}}
\newcommand{\tx}{\tilde{x}}
\newcommand{\ty}{\tilde{y}}
\newcommand{\om}{\omega}
\newcommand{\z}{\zeta}
\newcommand{\zb}{\bar{\zeta}}
\newcommand{\p}{{\bf p}}

\begin{document}

\title{Vortex structures of  rotating Bose-Einstein condensates in anisotropic harmonic potential }
\author{ S. I. Matveenko}
\affiliation{L.D. Landau Institute for Theoretical Physics, Kosygina Str. 2, 119334, Moscow, Russia}
\affiliation{LPTMS, CNRS and Universite Paris-Sud, UMR8626, Bat. 100, 91405 Orsay, France}

\date{\today}

\begin{abstract}
We found an analytical solution for the vortex structure in a rapidly rotating
trapped Bose-Einstein
condensate  in the lowest Landau level approximation.
 This solution is exact in the limit of a large number of
vortices and is obtained for the case  of anisotropic harmonic potential.
 For the case of symmetric harmonic trap when the rotation frequency is equal to the trapping
 frequency, the solution coincides with the Abrikosov  triangle vortex lattice in type-II superconductors.
 In a general case the coarse grained density is  found to be close to the Thomas-Fermi profile,
  except the vicinity of edges of a  condensate cloud.

\end{abstract}

\maketitle

Bose-Einstein condensates (BEC)  are a new state of matter where various aspects of macroscopic
quantum physics can be studied.
The experimental studying  of BEC in ultra-cold  rotating atomic gases shows
  a wide variety of new features in the physics of quantized vortices and vortex
arrays \cite{exp,coop,revfet} which can be not accessible in other quantum systems containing vortices,
such as superfluid   helium
or type-II superconductors.
 In a harmonically trapped condensate
rotating at a frequency close to the trap frequency,
 vortices form a triangular Abrikosov lattice
 and the coarse grained density approaches an inverted
parabola \cite{wat,num2,aft2,us}.  At very fast rotation, when the number
of vortices becomes close to the number of atoms,
the states are strongly correlated and the vortex lattice
is expected to melt.

We propose below the solution for the vortex structure
of rotating BEC in anisotropic harmonic trapping. The found function satisfiers the
projected to the lowest Landau level (LLL) Gross-Pitaevskii
(GP) equation with extremely high accuracy for  a large number of vortices.

We consider a system of bosonic  atoms strongly confined in the $z$ direction by an external trapping
potential  such that the bosons   become essentially two-dimensional. The bosons are confined in the
plane by a anisotropic harmonic trapping potential $V(\rr) =m(\om_x^2 x^2+\om_y^2 y^2)/2$, with
$\om_y<\om_x$, for a definiteness, and the trap is rotating around the $z$ axis with frequency
$\Omega$. We assume  that all
particles are in the same macroscopic quantum state described by the wave function $\psi({\bf r})$.
 In the rotating frame the Gross-Pitaevskii equation for $\psi({\bf r})$ reads:
\beq
\frac{\hat{\p}^2}{2 m} \psi + g |\psi|^2 \psi  +V(\rr) \psi  - \Omega \hat{L}_z \psi = \mu \psi,
\label{gp}
\eeq
where $\hat{\p}$ is the momentum operator, $m$ is the particle mass,
 $g > 0$ is an effective 2D coupling constant,
 $\hat{L}_z$ is the operator of the orbital angular momentum, $\mu$ is the chemical potential, and $\psi$ is
normalized to the total number of particles $N$.

 As it is well known, the single-particle Hamiltonian $H_0$ for rotating neutral atoms is equivalent to the
  Hamiltonian of a charged particle in a uniform magnetic field $B= 2 m \Omega$ along the $z$ axis.
   The half the cyclotron frequency is equal to  $\omega_c= \Omega$, and the
vector-potential in the symmetric gauge is ${\bf{A}}={\bf B}\times\rr/2=m{\bf{\Omega}}\times\rr$.
\beq
\begin{split}
&H_0=\frac{\hat{\p}^2}{2 m} - \Omega \hat{L}_z  + V(\rr) =
\frac{1}{2 m}{(\hat{\p} -  {\bf{A}} /c)^2} \\
&+ \frac{1}{2}m (\om^2_x -\Omega^2)  x^2+  \frac{1}{2}m (\om^2_y -\Omega^2)  y^2.
\end{split}
\label{Hsingle}
\eeq

Below the critical rotation frequency, $\Omega<\om_x, \, \om_y$,  the
presence of the residual confining potential lifts the  degeneracy of the Landau levels.
Provided that $\om_x -\Omega, \, \om_y - \Omega \ll \Omega $ and interactions are mach smaller than the cyclotron frequency
($n g \ll 2 \Omega $, where $n$ is  the two-dimensional particle density), we can restrict our consideration
 to the lowest Landau level.

The Gross-Pitaevskii equation projected onto the LLL of an asymmetric harmonic trap can be written
in terms of dimensionless variables $\tx$, $\ty$ as \cite{us}
\beqa
\label{pgp}
&(\mu-\omega_t^{+})f(\z) = \frac{\omega_t^{+}-\omega_t^{-}}{ 2}
(- \sinh (2 \nu ) f''(\z) \nonumber \\
&+ 2 \z f'(\z))+
 g\frac{1}{ \pi} \text{e}^{-\tanh(\nu) \z^2/2} \nonumber \\
&\times \int d\z' d\zb'\text{e}^{-2 \z'\zb' +\z \zb' + \tanh(\nu) \z'^2 + \tanh(\nu)\zb'^2 /2    }  f^2(\z')\overline{f(\z')} .
\eeqa
where
\beq
\z = \tx + i \ty = \frac{i}{2} \sqrt{\om_t^+ -\om_t^-} \sqrt{\sinh 2 \nu} \left[ z \rho - \frac{\bar{z}}{\rho}\right], \quad z = x + i y,
\eeq
\beq
\tanh \nu = \frac{\om_t^+}{\om_t^-}\sqrt{\frac{{\omega_t^-}^2 -\omega_c^2}{{\omega_t^+}^2 -\omega_c^2}},
 \qquad \rho^2=\sqrt{\frac{(\omega_t^{-}+\omega_c)(\omega_t^{+}+\omega_c)}
 {(\omega_t^{-}-\omega_c)(\omega_t^{+}-\omega_c) }},
\eeq
with
 $\tilde\omega_x^2=\omega_x^2-\omega_c^2$,
$\tilde\omega_y^2=\omega_y^2-\omega_c^2$,
 $\omega_t^{\pm}=\sqrt{\omega_c^2+(\frac{\tilde\omega_x\pm \tilde\omega_y}{2})^2}$,
and $d \tx \, d \ty = (\om_t^+ \om_t^- /\om_c ) dx \, dy$.
The LLL wave functions have a form
\beq
\Psi (x, y) = f(\z ) \text{e}^{-\frac{1}{2}\omega_t^{+}z\bar z}e^{-\frac{1}{2}(az^2+b\bar z^2)},
\eeq
with  an analytic function $f(\z)$, and  $ a=\frac{1}{ 2}\rho^2(\omega_t^{-}-\omega_c)$,
 $ b=\frac{1}{2}(\omega_t^{-}+\omega_c)/{ \rho^2}$.

The  anisotropy parameter  $\tanh \nu$ is equal to zero  for the cylinder symmetric trap: $\om_x = \om_y \geq \Omega$.
 The opposite quasi-one-dimensional case with a narrow channel geometry
 is achieved  at $\tanh\nu  \to 1$, when the rotation frequency becomes equal to the
 lowest trapping frequency: $\Omega  \to \omega_y < \omega_x $. Both limited  cases were considered in
  details in \cite{us}.

Note that for the case  of $\omega_x = \omega_y =\Omega$ the free energy
 $\int dx \, dy [\psi^* H_0 \psi + g|\psi|^4 /2 -\mu |\psi|^2]$
has the same form as   the Ginzburg-Landau functional of a superconductor in the magnetic field
 near  a phase transition. The important
 difference and  simplification in our case is the constant "magnetic field"  $2 m \Omega$ (infinite penetration depth).
  Therefore, instead of  three equations  obtained by variation over $\psi$, $B$,  plus the boundary condition,  we have one equation
 with the normalization condition $\int dx \, dy |\psi^2| = N$.
   Since the penetration
  depth of the "magnetic field"  is infinity,  properties of our system will be similar
  to   properties of type-II superconductors.
  Therefore it is reasonable to expect that superconductors in the magnetic field and rotating BEC will
   have similar structure of vortex lattice.
  Indeed, we can see that  the well known approximate solution
 for Abrikosov vortex lattice \cite{abr} was built  by use of
   the lowest Landau level (LLL) wave functions, as well as the solution for the BEC.
 Moreover, an approximate solution for Abrikosov lattice in type-II superconductor
 becomes the exact solution for the
  considered   model of rotating BEC in the LLL approximation.

For the general case of asymmetric harmonic potential the ground state
 contains an ordered  vortex structure in the parameters region
 $Ng/l^2 /(\hbar(\omega_{x,y}-\Omega)) \gg1$, where
$ l=(\hbar/m\Omega)^{1/2}$ is  the effective magnetic length.
The number of vortices will increase with the increase of this ratio. When the  number
of particles  will become of the order  of the number of vortices, the  vortex lattice will melt through
 the phase transition to a strongly correlated state. In this region the Gross-Pitaevskii  mean field equation is not applicable.

 We will  find the solution for the  vortex structure
as a  special  deformation of the exact solution for spatially homogeneous vortex lattice
for the case of a cylinder symmetric potential ($\omega_x =\omega_y = \omega$) at the critical value of the
frequency rotation $\Omega = \omega$:
\begin{equation}
f_0(\z) = \frac{(2v)^{1/4}}{\sqrt{S}}\vartheta_1 \left(\pi \z /b_1, \tau \right)\exp(\pi \z^2/2v_c),
\label{f}
\end{equation}
where $S$ is the surface area,  and $\vartheta_1$ is the Jacobi theta-function  given by
\begin{equation}
\label{theta}
\vartheta_1 ( \eta, \tau )=\frac{1}{i}\sum_{n= -\infty}^{\infty}(-1)^{n}\exp\{i\pi\tau(n + 1/2)^2+2i\eta (n+1/2)\}.
\end{equation}
Real ($u$) and imaginary     ($ i v$) parts  of the   quasiperiod $\tau=u+iv$  will be fixed below.
The  Jacobi theta-functions are analytic in the  complex plane and have zeros at the points $\eta =n \pi + m \pi \tau$,
where $n,m $ are integers. The function $f_0 (\z)$ has zeros at the lattice sites $n b_1 + m b_2$, with $b_2 = b_1 \tau$. These points correspond to the vortex
locations.
The parameter $b_1$ can  be chosen real so that  the area  of the unit cell is $v_c = b_1^2 v$.
Using the property  of the $\vartheta$-function:
\[
| \vartheta_1 \left(\pi \z /b_1, \tau \right)| = G(\tx, \ty) \exp\left[\frac{\pi \ty^2}{v_c}\right],
\]
with a periodic oscillating $G$: $0 \leq G(\tx, \ty) \leq \vartheta_2 \left(0, \tau \right)$, one can
 obtain that  the envelope  $\overline{|f_0(\z)|}$  has a polar symmetric form
\beq
\overline{|f_0(\z)|}\sim \exp \left[\frac{\pi(\tx^2 + \ty^2)}{2 v_c} \right].
\label{polarsymmetric}
\eeq
The function $f_0(z)$  with  fixed elementary cell area  ${v_c} = \pi$ is an exact solution of Eq.~(\ref{pgp}) for the case of a critical rotation $\omega_x =\omega_y =\Omega$.
The normalization coefficient in Eq.~(\ref{f}) is chosen such that the function
 $\Psi =[f_0(z)/l]\exp(-z{\bar z}/2)$ is normalized to unity. The function $\Psi$ has  a constant
 envelope and describes
a periodic vortex structure. The minimum energy is obtained for the
 triangular lattice, where $\tau = \exp 2\pi i/3$, $v=\sqrt{3}/2$, and $b_1^2 = 2 \pi/\sqrt{3}$. The chemical
potential is then given by
 $\mu=\frac{\alpha Ng}{l^2}$,
with $\alpha=(3^{1/4}/2)\sum_{b,c}(-1)^{mp}\exp\{-\pi^2(b^2+c^2)/4b_1^2\}=1.1596$, and $b=2m$, $c=2p$
 being even integers.

In the general case ( $\Omega < \omega_x, \omega_y$), we will find a  solution of equation
 (\ref{pgp})  in the form
\beq
f(\z) = (2v)^{1/4}\sum_{n= -\infty}^{\infty} (-1)^n  \hat{g}(a) \tilde{q}^{a^2}
\text{e}^{i \sqrt{\pi v}a \z+ \frac{\z^2}{2}(1 - \tanh \nu )},
\label{symf}
\eeq
where  $a=2n+1$ are odd integers, $\tilde{q} = \exp[i \pi \tau /4]$, and $\hat{g}(a)$ is a  differential operator acting on  $a$.  Substituting the trial function (\ref{symf}) into equation (\ref{pgp}) for a triangular-like lattice (the lattice that becomes exactly triangular for $\omega=\Omega$) we
obtain:
\begin{equation}
\begin{split}
 \biggl\{\left(\mu -\om_t^+ +  (\om_t^+ - \om_t^-)\left[ \frac{\tanh(\nu) }{1 + \tanh (\nu)}  -
\hat{A}  \hat{A}^+ \right] \right)\hat{g}(a)     \\
-\sqrt{v} g \sum_{b, c} \hat{g}(a-b)\hat{g}(a-c)\overline{\hat{g}(a - b - c)} \\
\times \exp\left[-\frac{\pi v}{4 }(b^2
+ c^2)\right](-1)^{m p}  \biggr\}
 \tilde{q}^{a^2}\exp\left[i \sqrt{\pi v} a \z\right]=0.
\end{split}
\label{mainsym}
\end{equation}
Here  we introduced the operators $\hat{A}, \hat{A}^{\dagger}$ which a creation (annihilation) operators for
a corresponding harmonic oscillator with usual commutation relations $[\hat{A}, \hat{A}^{\dagger}] = 1$:
 \[
 \hat{A} , \hat{A}^{\dagger} =  \frac{ \sqrt{\pi \tilde{v}}}{2} a    \pm \frac{1}{\sqrt{\pi \tilde{v}}} \frac{\partial}{\partial a},
 \]
 where $\tilde{v} = v \gamma$, with $\gamma = (1+\tanh \nu)/(1 - \tanh \nu)$.

 For large $\beta \equiv N g /(l^2 (\omega_t^+ - \omega_t^-)) \gg1$
  and  $\mu^*  \equiv (\mu - \om_t^+) /(\om_t^+ -\om_t^-) -1/(1 +\tanh \nu)   \gg 1$  an approximate
   solution for $\hat g(a)$, which describes the vortex structure with a high accuracy, has the form
\beq\label{energybis}
\hat{g}(a) = \frac{1}{\sqrt{\alpha\beta}}  \sqrt{R^2 - \hat{A}^{\dagger} \hat{A}}\,\,\Theta(R^2-    \hat A^{\dagger}\hat A),
\eeq
where $\Theta$ is the Heaviside step function, $R=\sqrt{\mu^*}$, and $\alpha = 1.1596$.
We will see below that  for the symmetric case  $R$  is the radius of the condensate cloud in units of $l$.
Equation (\ref{energybis}) is obtained taking into account that the leading contribution to the sum over
 $b$ and $c$ in Eq.~(\ref{mainsym}) comes from small values of $m$ and $p$,
since already the contributions of terms with $|m|\geq 2$ or $|p|\geq 2$ are exponentially small.
 Provided that the dependence  $\hat{g}(a)$ is smooth, which is the case for large
$R$,  we may consider large $a$ and omit $b$ and $c$ in the arguments of the $\hat g$ operators in
 Eq.~(\ref{mainsym}).  This immediately gives Eq.~(\ref{energybis}). Mathematically, the high accuracy
of the solution is based on a known fast convergency of the series for $\vartheta$-function due to
the multiplier  $\tilde{q}^{(2 n + 1)^2}$.

Substituting the solution (\ref{energybis}) into Eq.(\ref{symf})
and expanding (\ref{symf})  into series in  known eigenfunctions of harmonic oscillator
we obtain after some algebra:
\beq
\begin{split}
&f(\z) = \frac{(2v)^{1/4}\sqrt{(1 + \tanh \nu)} }{\sqrt{\alpha\beta}} \sum_{n = -\infty}^{\infty} \sum_{k = 0}^{[R^2]} (-1)^{[n(n-1)/2]}\, \\
&\times\sqrt{R^2 -  k}\,\frac{ (i \sqrt{\tanh \nu})^k}{2^{k} k!}\\
& \times H_k\left(\frac{\z}{\sqrt{\sinh 2\nu}}\right)
H_k\left(\sqrt{\frac{\pi \tilde{v}}
{2}} a\right)  \text{e}^{-\pi \tilde{v} a^2 /4},
\end{split}
\label{ffin}
\eeq
where $H_k(w)$ are Hermite polynomials, $a = 2 n + 1$.  The solution (\ref{ffin}) is simplified
 in the symmetric case, where $\nu \to 0$, $H_k(\z/\sqrt{\sinh 2\nu}) \to 2^{k/2} \z^k /\nu^{k/2}$,
 and  in the one-dimensional case where $\nu \to \infty$ and operators $A$, $A^{\dagger}$  becomes
  numbers: $A=A^{\dagger}   \propto a$.

\begin{figure}{}
\includegraphics[width=3.0in]{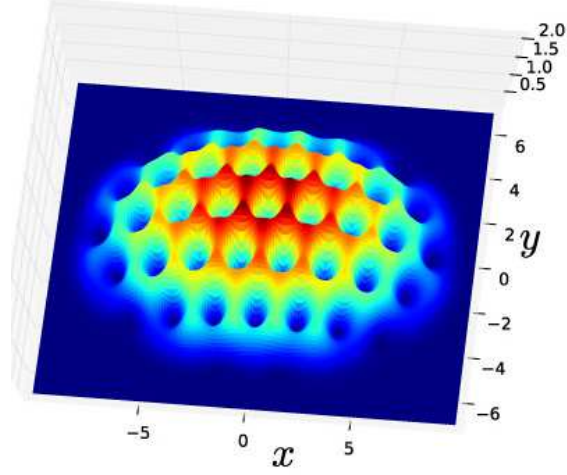}
\caption{(Color
  online)  Condensate wave-function $|\psi(x,y)|^2$ for $ R=7 $, $\tanh \nu = 1/4$. Coordinates $x$
 and $y$  are given in units of $l$.}
\label{fign}
\end{figure}

From the condition that the function $[f(\z)/l]\exp(-\z{\bar \z})$ is normalized to unity we find a relation
$R=(2\alpha\beta \gamma/\pi)^{1/4}$.  For the symmetric potential  ($\gamma = 1$) this result
 is  in agreement with  Refs. \cite{num2,aft2,aft1,she,baum} .
As mentioned above the solution (\ref{ffin}) includes limiting cases of cylinder and
narrow channel geometries, considered
in details in \cite{us}.  Numerical results \cite{us} for these cases demonstrated excellent coincidence with the analytical solution.
The structure of the vortex lattice for $R=7$, $\tanh\nu = 1/4$ is shown in Fig.~\ref{fign}.
Averaging the density over the oscillations,
that is averaging  the density $|\Psi|^2$ over a distance scale much larger than $l$, gives the coarse grained density:
\beq
\begin{split}
&\bar{n}_{cg}(r)=\frac{n_{cg}(0)}{\cosh \nu} \sum_{k=0}^{[R^2]}\left (1 - \frac{k}{R^2}\right)\,   \frac{ (\tanh \nu)^{k}}{ 2^k k!} H_k\left(\frac{\z}{\sqrt{\sinh 2\nu}}\right) \\
&\times H_k\left(\frac{\bar{\z}}{\sqrt{\sinh 2\nu}}\right)
\text{e}^{-|\z|^2 + (\z^2 +\bar{\z}^2)\tanh(\nu) /2}  \\
& =\frac{n_{cg}(0)}{\pi \cosh \nu} \int\int_{-\infty}^{\infty} dt_1 dt_2  \text{e}^{-t_1^2 -t_2^2} \sum_{k=0}^{[R^2]}\left (1 - \frac{k}{R^2}\right)\,
 \frac{X^{k}}{  k!}  \\
&\times \text{e}^{-|\z|^2 + (\z^2 +\bar{\z}^2)\tanh(\nu) /2}
= \frac{n_{cg}(0)}{\pi \cosh \nu} \int\int_{-\infty}^{\infty} dt_1 dt_2  \text{e}^{-t_1^2 -t_2^2 + X}\\
&\times \frac{\Gamma(1 + R^2, X) - X \Gamma(R^2, X)}{\Gamma(R^2 + 1)}
 \text{e}^{-|\z|^2 + (\z^2 +\bar{\z}^2)\tanh(\nu) /2},
\end{split}
\label{cgd}
\eeq
where $X(t_1, t_2 ) = 2\tanh( \nu) (-i \z /\sqrt{\sinh(2 \nu)} +  t_1)(i\bar{\z}/\sqrt{\sinh(2 \nu)} + t_2 )$.
In the limit $\nu  \to 0$  when $X \to |\z|^2$, we obtain exactly  the expression for the symmetric
 case \cite{us}.  The  solution for a narrow channel case  is obtained   in the limit   $\nu \to \infty$
 taking into account the divergence  of $R$: $R \propto (1 - \tanh \nu )^{-1/2}$.

The expression (\ref{cgd}) can be simplified in the case of large $R$. Inside the condensate cloud we
obtain the density profile which is  close to the Tomas-Fermi inverted paraboloid:
\beq
\begin{split}
&n = n_{cg}(0)\bigg[1 - \frac{1}{R^2 (1-\tanh^2\nu)}[\tanh^2\nu  \\
& +\tilde{x}^2(1 -\tanh\nu)^2 + \tilde{y}^2 (1 + \tanh \nu)^2]\bigg],
\end{split}
\eeq
with $n_{cg}(0) \approx 2N/(\pi R^2)$.
The condensate is localized at the region inside the ellipse
$x^2/a^2  + y^2/b^2 =1$ with  semiaxes  $a \geq b$:

\[
 a \approx {R} \sqrt{\gamma},  \qquad b\approx  \frac{R}{\sqrt{\gamma}}.
\]
Outside the ellipse the density  is  negligible  small  and has the  asymptotic form
\beq
\begin{split}
n_{cg} \propto \exp[R^2 \ln(|\z |^2 (1 - \tanh \nu )^2) \\
-\tilde{x}^2 (1-\tanh\nu) -\tilde{y}^2(1 + \tanh\nu)].
\end{split}
\eeq
 Outside the condensate cloud  the density drops exponentially at distance of the order of magnetic length
 or intervortex spacing.
 In the direction $y$, for example,  in the interval
($1 \ll y - b \ll b^{1/3}$) we obtain
\beq
n_{cg} \sim  {n_{cg}(0)} \frac{\text{e}^{- 2(\tilde{y}- b)^2}}{4 \sqrt{2 \pi} b (y - b)^2}.
\eeq

\begin{figure}[htb]
 \includegraphics[width=3.0in]{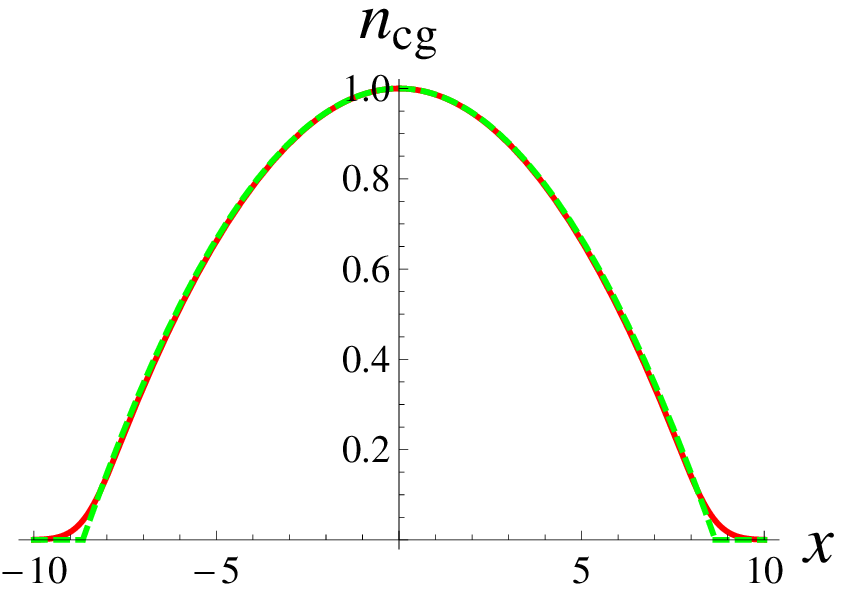}
 \caption{(Color
  online) The coarse grained density $n_{cg} (x, 0)$ in units of ${\bar n}_{2D}$ versus $x$, for $R=5$, $\tan \nu = 1/2$.
 The dashed curve shows the Thomas-Fermi
inverted-parabola
shape, and $x$ is given in units of $l$.}
 \label{figx}
 \end{figure}
\begin{figure}[htb]
 \includegraphics[width=3.0in]{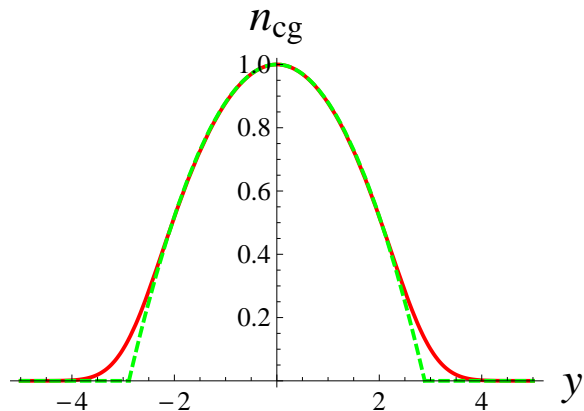}
 \caption{(Color
  online) The coarse grained density $n_{cg} (0, y)$ in units of ${\bar n}_{2D}$ versus $y$,
 for $R=5$, $\tan \nu = 1/2$. The dashed curve shows the Thomas-Fermi
inverted-parabola
shape, and $y$ is given in units of $l$.}
 \label{figy}
 \end{figure}

At the Thomas-Fermi border we have ${ n}_{cg} \sim {n_{cg}(0)}/\sqrt{2\pi |\z|^2}$.
Our results show that inside the condensate cloud  the density profile has   the Thomas-Fermi
 inverted-paraboloid  shape, except the  region  close to the edge:
  $|\delta z | \lesssim l$, where $|\delta z|$ is the distance to the border.
  We see in
   Figs.~\ref{figx},~\ref{figy}  that for $R=5$, $\tan \nu = 1/2$ the Thomas-Fermi formula works well
    already for $|\delta \z |  \gtrsim 1 $.
The parabolic coarse grained density profile for the asymmetric potential was considered analytically and
 numerically in \cite{aft3,fet,oktel}.
Deviations from the Thomas-Fermi density profile of ${\bar n}_{cg}(r)$ have been studied in
 Ref. \cite{num2} for the symmetric  potential by using the variational numeric methods.
 Here we presented an analytic solution for  the vortex lattice  in a rapidly  rotating BEC in an asymmetric
  harmonic trap. The found coarse grained density profile is close to inverted paraboloid form.
The solution is  asymptotically exact in the  limit of  a  large number
of vertices.

\begin{acknowledgments}
I would like to thank S. Ouvry and G. Shlyapnikov for useful discussions.
\end{acknowledgments}

\end{document}